\documentclass[cits]{PoS}
\usepackage{colordvi}
\usepackage{bbm}

\newcommand{\be}{\begin{equation}}
\newcommand{\ee}{\end{equation}}
\newcommand{\bea}{\begin{eqnarray}} 
\newcommand{\eea}{\end{eqnarray}}
\newcommand{\gtilde}{\frac{g^2}{16 \, \pi^2}\; }
\newcommand{\qqq}{\displaystyle{\not}q_d}
\newcommand{\qq}{\displaystyle{\not}q_s}
\newcommand{\bpsi}{\overline{\psi}}
\newcommand{\Dr}{\displaystyle{\not}{\overrightarrow{D}}}
\newcommand{\Dl}{\displaystyle{\not}{\overleftarrow{D}}}
\newcommand{\partl}{\displaystyle{\not}{\overleftarrow{\partial}}}
\newcommand{\partr}{\displaystyle{\not}{\overrightarrow{\partial}}}
\newcommand{\DDr}{\overrightarrow{D}}
\newcommand{\DDl}{\overleftarrow{D}}
\newcommand{\MSbar}{{\overline{\rm MS}}}

\title{The chromomagnetic operator on the lattice}

\ShortTitle{The chromomagnetic operator on the lattice}

\bigskip
\author{M. Constantinou$^a$, M. Costa$^a$, R. Frezzotti$^b$,
  V. Lubicz$^c$,
 G. Martinelli$^d$, D. Meloni$^c$,
  \speaker{H.~Panagopoulos}\,$^a$, S. Simula$^c$ \\
\llap{$^a$} Department of Physics, University of Cyprus, CY-1678 Nicosia,
  Cyprus\\
\llap{$^b$} Dipartimento di Fisica, Universit\`a di Roma ``Tor Vergata''
and INFN Sezione ``Tor Vergata'', I-00133 Rome, Italy\\
\llap{$^c$} Dipartimento di Fisica, Universit\`a Roma Tre, and INFN,
Sezione di Roma Tre, I-00146 Rome, Italy\\
\llap{$^d$} SISSA, I-34136 Trieste, Italy

\bigskip
E-mail: \email{constantinou.martha@ucy.ac.cy},
        \email{kosta.marios@ucy.ac.cy}, \email{roberto.frezzotti@roma2.infn.it},
        \email{lubicz@fis.uniroma3.it}, \email{guido.martinelli@sissa.it},
        \email{meloni@fis.uniroma3.it}, \email{haris@ucy.ac.cy}, \email{simula@roma3.infn.it}}

\abstract{We study matrix elements of the ``chromomagnetic'' operator on the
lattice. This operator is contained in the strangeness-changing
effective Hamiltonian which describes electroweak effects in the
Standard Model and beyond.

Having dimension 5, the chromomagnetic operator is characterized by a
rich pattern of mixing with other operators of equal and lower
dimensionality, including also non gauge invariant quantities; it is
thus quite a challenge to extract from lattice simulations a clear
signal for the hadronic matrix elements of this operator.

We compute all relevant mixing coefficients to one loop in lattice
perturbation theory; this necessitates calculating both 2-point
(quark-antiquark) and 3-point (gluon-quark-antiquark) Green's functions
at nonzero quark masses. We use the twisted mass lattice formulation,
with Symanzik improved gluon action.

For a comprehensive presentation of our results, along with detailed
explanations and a more complete list of references, we refer to our
forthcoming publication \cite{CMO}.}

\FullConference{31st International Symposium on Lattice Field Theory LATTICE 2013\\
		 July 29 -- August 3, 2013\\
		 Mainz, Germany}

\begin{document}

\section{Introduction}

The electroweak effective Hamiltonian describing stangeness changing
($\Delta S =1$) processes, in the Standard Model (SM) and beyond, contains 
four ``magnetic'' operators of dimension 5:
\be H^{\Delta S = 1,\ d=5}_{\rm eff} = \sum_{i=\pm} (C^i_\gamma Q^i_\gamma + C^i_g Q^i_g) + {\rm h.c.}\ee
\vspace{-0.3cm}
$$ Q_\gamma^\pm = {Q_d\,e\over 16 \pi^2} \left(\bar s_L \,\sigma^{\mu\nu}\, F_{\mu\nu} \, d_R 
\pm \bar s_R \,\sigma^{\mu\nu}\, F_{\mu\nu} \, d_L \right),\quad
Q_g^\pm = { g\over 16 \pi^2} \left(\bar s_L \,\sigma^{\mu\nu}\,
G_{\mu\nu} \, d_R  
\pm \bar s_R \,\sigma^{\mu\nu}\, G_{\mu\nu} \, d_L \right)$$
The coefficients $C^i_{\gamma}$ and $C^i_{g}$, multiplying the
electromagnetic (EMO) and chromomagnetic (CMO) operators,
respectively, may be calculated perturbatively via the
OPE; they are suppressed within the
SM, but become more pronounced beyond the SM, e.g.
through penguin diagrams in SUSY.

The matrix elements of the CMO are parameterized as \cite{DIM}:
\bea \displaystyle\langle\pi^0|Q_g^+|K^0\rangle &=& {-11\over
  32\sqrt{2}\pi^2}\, {M_K^2 (p_\pi\cdot p_K)\over m_s+m_d}\, \NavyBlue{B_{g1}}\label{piQK}\\
\displaystyle\langle\pi^+\pi^-|Q_g^-|K^0\rangle &=&
        {11\,{\rm i}\over
  32\pi^2}\, {M_K^2\,M_\pi^2\over f_\pi\,( m_s+m_d)}\,  \NavyBlue{B_{g2}}\label{pipiQK}\\
\displaystyle\langle\pi^+\pi^+\pi^-|Q_g^+|K^+\rangle &=& {-11\over
  16\pi^2}\, {M_K^2\,M_\pi^2\over f_\pi^2\,( m_s+m_d)}\, \NavyBlue{B_{g3}}\label{pipipiQK}
\eea
These matrix elements are relevant for the study of $K^0-\bar K^0$
mixing, $\epsilon^\prime/\epsilon$, the $\Delta I = 1/2$ rule, and
$K\to 3\pi$ decays.
To leading order in $\chi$PT, the \NavyBlue{$B$}-parameters are all
related \cite{BEF}:
\be
\displaystyle Q_g^\pm = {11\over 256\pi^2}\, {f_\pi^2\,M_K^2 \over m_s+m_d}\,
\NavyBlue{B_{g}}\,\left[U(D_\mu U^\dagger)(D^\mu U) \pm (D_\mu
  U^\dagger)(D^\mu U) U^\dagger \right]_{23} \ee
Thus, a lattice study of, say, Eq.~(\ref{piQK}),
provides information for Eqs.~(\ref{pipiQK}), (\ref{pipipiQK}) as well.

The EMO has been studied in simulations with $N_f = 0$~\cite{BLMM} and
$N_f=2$~\cite{BLMOS} dynamical flavors, focusing on: 
\be \displaystyle \langle\pi^0|Q_\gamma^+|K^0\rangle = {\rm i}\,
  {Q_d\, e\, \sqrt{2}\over 16\pi^2 \, M_K}\, p_\pi^\mu\, p_K^\nu \,
  F_{\mu\nu} \, \NavyBlue{B_T} \, R_T(q^2) \qquad [R_T(0) = 1]\ee

The parameter $\NavyBlue{B_T}$ appears, e.g., in the branching ratio
of $K_L \to \pi^0\,e^+\,e^-$ in SUSY models. 

\section{Operator Mixing -- Lattice Action -- Symmetries}

A formidable issue in the study of the CMO is the fact that it mixes
with a large number of other operators under renormalization. 
Even in dimensional regularization (DR), which has the simplest mixing
pattern, the CMO (${\cal O}_{CM}\equiv {\cal O}_1$) mixes with a total
of 9 other operators (${\cal O}_2-{\cal O}_{10}$), forming a basis of
dimension-five, Lorentz scalar 
operators with the same flavor content as the CMO. Among them, there
are also gauge noninvariant operators (${\cal O}_9\,,{\cal O}_{10}$);
these are BRST invariant and vanish by the equations of motion, as required by
renormalization theory.

\bea
\begin{array}{ll}
\Red{{\cal O}_1}    = g\,\bpsi_s \sigma_{\mu \nu} G_{\mu \nu}\psi_d  &
\Red{{\cal O}_6}    = \bpsi_s(\Dr+m_d)^2\psi_d + \bpsi_s(-\Dl +m_s)^2\psi_d \\
\Red{{\cal O}_2}    =  (m_{d}^2+m_{s}^2)\bpsi_s\psi_d &
\Red{{\cal O}_7}    = m_{s}\bpsi_s(\Dr+m_d)\psi_d + m_{d}\bpsi_s(-\Dl +m_s)\psi_d \\
\Red{{\cal O}_3}    =  m_{d}\,m_{s}\bpsi_s\psi_d &
\Red{{\cal O}_8}    = m_{d}\bpsi_s(\Dr+m_d)\psi_d + m_{s}\bpsi_s(-\Dl +m_s)\psi_d \\
\Red{{\cal O}_4}    =  \bpsi_s \DDl_\mu \DDr _\mu \psi_d &
\Red{{\cal O}_9}    =  \bpsi_s \Red{\partl}(\Dr+m_d)\psi_d 
                 - \bpsi_s (-\Dl +m_s)\Red{\partr} \psi_d \\
\Red{{\cal O}_5}    =  \bpsi_s(-\Dl +m_s)(\Dr+m_d)\psi_d \qquad&
\Red{{\cal O}_{10}}  = \bpsi_s \Red{\partr} (\Dr+m_d)\psi_d
                  - \bpsi_s (-\Dl +m_s)\Red{\partl}\psi_d
\end{array}
\eea

On the lattice, the mixing pattern can become considerably more
complicated, given that certain symmetries are violated; there can be
mixing with additional operators of dimension five (with finite
coefficients) or less (with power-divergent coefficients). A generic
hypercubic- and gauge-invariant lattice discretization will result in
mixing with 2+8+32 candidate operators of dimension 3, 4, 5, respectively. It is
thus imperative to make a judicious choice of lattice action, with a
large set of discrete symmetries, so as to exclude as many as possible
of these candidates.

We have adopted the twisted mass action for valence quarks and the
Osterwalder - Seiler action for sea quarks~\cite{FR} (along with a
compensating ghost action for valence quarks). For our one-loop 
perturbative calculation we only need the valence quark action, which
reads (in the physical basis):
\be S_F[\psi_f,\bar \psi_f,U]= a^4 \sum_f\,\sum_x\,\bar \psi_f(x) \Big{[}\gamma\cdot\widetilde\nabla
-i\gamma_5 W_{\rm{cr}}(r_f)+m_f\Big{]}\psi_f(x)\ee
\vspace{-0.2cm}
$$\gamma\cdot\widetilde\nabla\equiv\frac{1}{2}\sum_\mu\gamma_\mu
(\nabla^\star_\mu+\nabla_\mu) \qquad
W_{\rm{cr}}(r_f)\equiv-a\frac{r_f}{2}\sum_\mu\nabla^\star_\mu\nabla_\mu+
M_{\rm{cr}}(r_f) 
$$
($r_f$: Wilson parameter for flavour $f=u,\,d,\,s$; 
$M_{\rm{cr}}(r_f) = - M_{\rm{cr}}(-r_f)$: corresponding critical mass).

For gluons we have used the Symanzik improved action; for our
perturbative results we employed several standard
choices of values for the Symanzik coefficients appearing
in that action~\cite{CMO}.

A number of discrete symmetries~\cite{FR} are present in our action; 
the CMO is invariant -- up to a possible minus sign -- under them, and
the same must then hold for all other 
operators which mix with the CMO. As a result, alongside the 10 operators
which mix in DR, only 3 additional ones appear on the lattice, and
they all have dimension less than five:
\be\Red{{\cal O}_{11}} = i\,r_d\,\bpsi_s\gamma_5(\Dr+m_d)\psi_d
+i\,r_s\,\bpsi_s(-\Dl +m_s)\gamma_5\psi_d \,,\ 
\Red{{\cal O}_{12}} = i\,(r_d\,m_{d}+r_s\,m_{s})\bpsi_s\gamma_5\psi_d \,,\ 
\Red{{\cal O}_{13}} = \bpsi_s\,\psi_d
\ee

\section{Renormalization Matrix}

Renormalized operators  ${\cal O}_i^R$ are related to bare
ones ${\cal O}_i$ via a $13{\times}13$ renormalization matrix $Z$:
\be {\cal O}_i=  \sum_{j=1}^{13} Z_{ij} {\cal O}_j^R
  \qquad ({\cal O}=Z {\cal O}^R,\ \ {\cal O}^R =Z^{-1} {\cal O})
\ee
The matrix elements $Z_{ij}$ depend both on the regularization $X$ ($X=$ $L$ (lattice), $DR$
(dimensional), etc.) and on the renormalization scheme $Y$ ($Y=$
 $\MSbar$, $RI'$, etc.); where confusion might arise, one should denote
them as $Z_{ij}^{X,Y}$. At tree level: $Z=\mathbbm{1}$. For ${\cal O}_1^R$,
we only need the first row of $Z^{-1}$ (and thus, to one loop, only the first row or $Z$: 
$Z_{1i}$). Clearly: $Z_{1,1}= 1 + {\cal O}(g^2)$, $Z_{1i} = {\cal
  O}(g^2)\ (i>1)$.  

To obtain $Z_{1i}$, we have calculated, to one loop and in an arbitrary
covariant gauge, the 2-point (quark-antiquark) and
3-point (quark-antiquark-gluon) bare amputated Green's functions of
${\cal O}_1$\,; these are related to the corresponding renormalized
Green's functions through:
\bea
\begin{array}{ll}
\langle\psi^R\,{\cal O}_1^R\,\bpsi^R \rangle_{{\rm amp}} =
Z_\psi\,\sum_{i=1}^{13}(Z^{-1})_{1i} \langle\psi\,{\cal O}_i\,\bpsi
\rangle _{{\rm amp}}\,, & \psi=Z_\psi^{1/2}\,\psi^R \\
\langle \psi^R\,{\cal O}_1^R\,\bpsi^R A_\nu^R \rangle_{{\rm amp}} =
Z_\psi\,Z_A^{1/2}\sum_{i=1}^{13}(Z^{-1})_{1i} \langle\psi\,{\cal
  O}_i\,\bpsi\,A_\nu \rangle _{{\rm amp}}\,,\qquad &  A_{\nu}=
Z_A^{1/2}\,A^{R}_{\nu}
\end{array}
\label{GF}\eea

The renormalization functions $Z_\psi$, $Z_A$ (as well as those for
the coupling constant ($Z_g$), the fermion mass ($Z_m$), and the ghost
field ($Z_c$)) were not all available for the actions considered in this
work, and had to be calculated as a prerequisite. We mention in passing
that $Z_\psi$ and $Z_m$ do not depend on flavor in mass-independent schemes.
We also note that both the 2-point and 3-point functions are necessary
in order to fix 
all $Z_{1i}$\,, but they are also sufficient.

[An alternative definition of the CMO: 
$\widetilde{\cal O}_{CM} \equiv  m\,{\cal O}_{CM}$
appears in the study of 4-fermi operators. In this case,
the renormalization matrix reads: $\tilde Z_{ij}=Z_m\,Z_{ij}$
  ($m^R \equiv Z^{-1}_m\,m$). Similarly, a factor of $Z_g$ must be
included in $Z_{ij}$, if Green's functions are computed using: 
$(1/g){\cal O}_{CM}$, rather: ${\cal O}_{CM}$\,.]

\begin{figure}[h!]
\parbox[b][6cm][c]{5cm}{\hspace{1cm}\includegraphics[height=0.7cm]{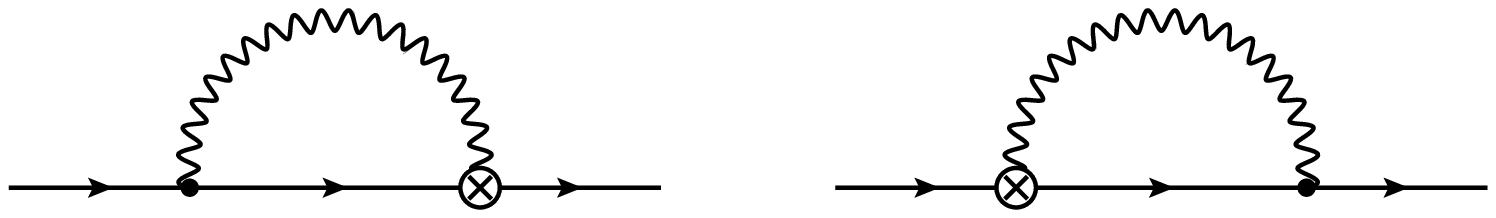}}\hspace{1cm}
\includegraphics[height=6.3cm]{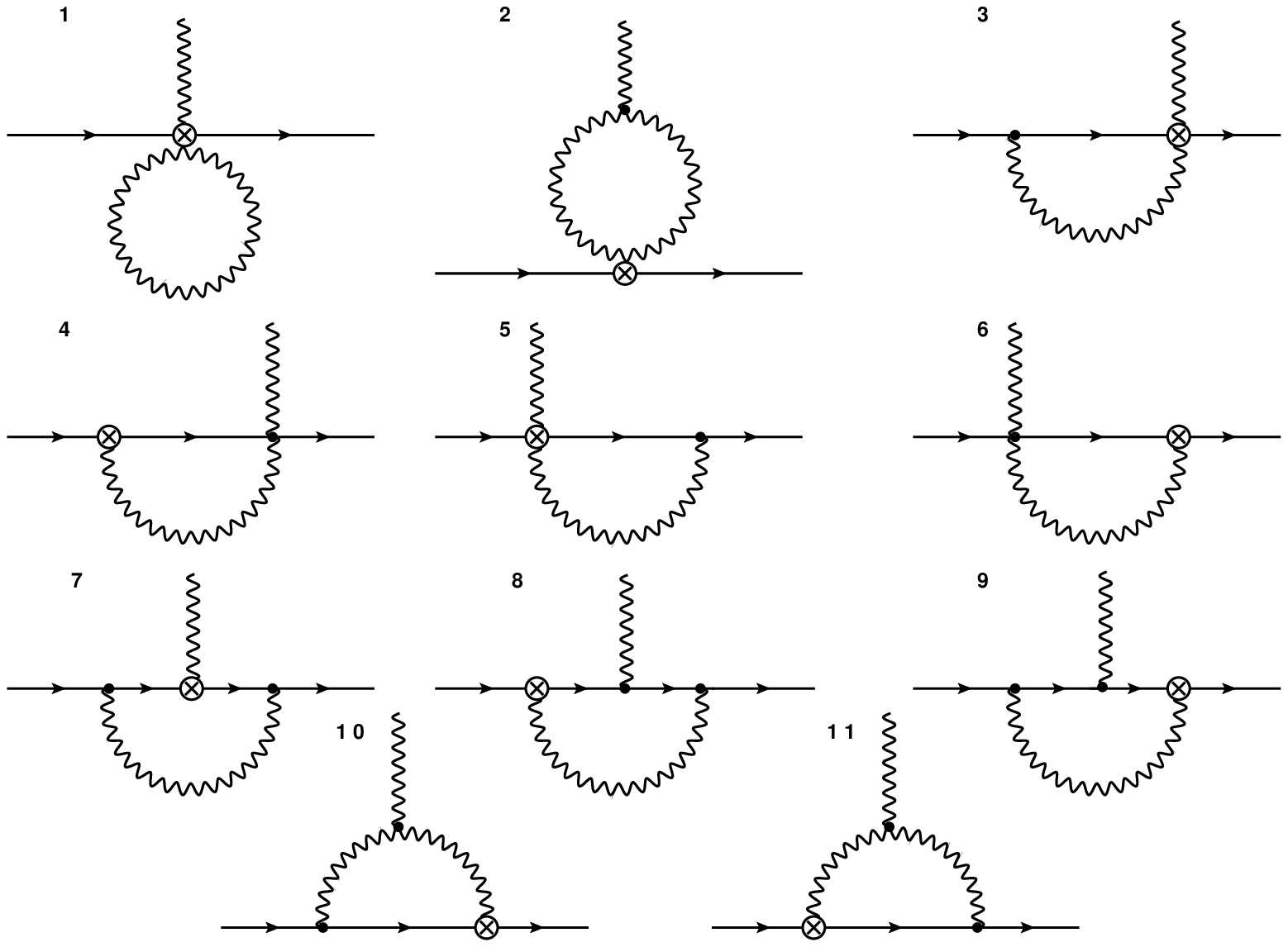}
\caption{1PI one-loop diagrams contributing to the 2-point and 3-point Green's
  functions. A wavy (solid) line represents gluons (quarks); an
  operator insertion is denoted by {\tiny${\otimes}$}.}
\label{2pt3pt}
\end{figure}

The one-particle irreducible (1PI) Feynman diagrams contributing to
the 2-point and 3-point Green's 
functions of ${\cal O}_1$ are shown in the left and right panels of Figure \ref{2pt3pt}, respectively. 

\section{Results}
We calculated the 2- and 3-point bare Green's functions of
Eq.~(\ref{GF}), first in DR and then in the far more complicated case
of the lattice. The purpose of the calculation in DR is twofold:
First, it provides the mixing coefficients $Z_{1i}^{DR,\MSbar}$, which are
interesting on their own right; second,
and most important, it leads to the renormalized Green's
functions in $\MSbar$, which are then necessary for extracting the
real quantities of interest: $Z_{1i}^{L,\MSbar}$.

\subsection{Dimensional Regularization and $\MSbar$ Renormalization}
In $D=4-2\epsilon$ dimensions, renormalizability requires that the
${\cal O}(1/\epsilon)$, 
1PI part in the bare Green's functions of Eq.~(\ref{GF}) has
polynomial dependence on $m_s$, $m_d$, $q_s$, $q_d$, $q_A$
($q_s/q_d/q_A$\,: momenta of the external antiquark/quark/gluon). In
fact, there appear in total 7+4 types of such dependence, as follows:
\bea
\langle\psi\,{\cal O}_1\,\bpsi
\rangle _{{\rm
    amp}}^{1-loop}\Bigr|_{1/\epsilon}\hspace{-0.1cm}&=&\hspace{-0.1cm}
\Red{\rho_1}\,(q^2_s {+} q^2_d) + \Red{\rho_2}\,(m_s^2 {+} m^2_d) +
\Red{\rho_3}\,i\,(m_d\qqq {+} m_s\qq) \nonumber \\ 
&&\hspace{-0.1cm}+ \Red{\rho_4}\,i\,(m_s\qqq {+} m_d\qq) +
\Red{\rho_5}\,q_s{\cdot}q_d + 
\Red{\rho_6}\,\qq\,\qqq + \Red{\rho_7}\,m_sm_d \nonumber \\
\langle\psi\,{\cal O}_1\,\bpsi \, A_\nu 
\rangle _{{\rm amp}}^{1-loop}\Bigr|_{1/\epsilon} &=& \Red{R_1}\,g
\,(q_s{+}q_d)_\nu + \Red{R_2}\,g\,(\gamma_\nu\qqq+ \qq\gamma_\nu)
\nonumber\\
&& +\Red{R_3}\,i\,g\,(m_s{+}m_d)\gamma_\nu + \Red{R_4}\,(-2
i\,g\,\sigma_{\rho\nu}q_{A\rho}) 
\eea
There exist
  also 1-particle reducible diagrams contributing to the 3-point
  function, both at tree level and at one loop; these contain
  non-polynomial ${\cal O}(1/\epsilon)$ terms, which however cancel by
  virtue of the 2-point relation.
Computing the coefficients $\rho_1 - \rho_7$, $R_1 - R_4$ we find:
\be  \rho_1 = \frac{g^2\,C_F}{16\,\pi^2}\,\frac{1}{\epsilon}(-3)\quad
\rho_2 = \frac{g^2\,C_F}{16\,\pi^2}\,\frac{1}{\epsilon}(-6)\quad
\rho_3 = \frac{g^2\,C_F}{16\,\pi^2}\,\frac{1}{\epsilon}(3)\quad
\rho_4 = \rho_5=\rho_6=\rho_7=0
\ee
\be \{R_1\,,\, R_2\,,\, R_3\,,\, R_4\} =
\frac{g^2}{16\,\pi^2}\,\frac{1}{\epsilon}\,\bigl\{-6\,C_F,\, 
\frac{3\,N_c}{4},\, 
(\frac{-3}{2\,N_c} +\frac{3\,N_c}{4} ),\, 
(\frac{1}{N_c}-\frac{\alpha}{2\,N_c}+\frac{7\,N_c}{4}
+\frac{3\,\alpha\,N_c}{4})\bigr\}
\ee
($N_c$\,: number of colors, $C_F = (N_c^2-1)/(2N_c)$, $\alpha$\,:
gauge parameter). 

Demanding that all ${\cal O}(1/\epsilon)$ dependence on the right-hand
sides of Eqs.~(\ref{GF}) disappears (as it ought to, since the
corresponding renormalized Green's functions on the left-hand sides
must be finite) provides 7+4 constraint equations on the 10 coefficients
$Z_{1i}$. This set of equations is self-consistent and complete; solving
them, we obtain:

\bea\begin{array}{ll} \displaystyle
Z^{DR,\MSbar}_{1,1}  = 1+
\gtilde\,\frac{1}{\epsilon}\,(-\frac{N_c}{2}+\frac{5}{2\,N_c})
\hspace{0.5cm} & \displaystyle
Z^{DR,\MSbar}_{1,2}  = -2\, Z^{DR,\MSbar}_{1,10} =
\gtilde\,\frac{1}{\epsilon}\,(-3\,N_c+\frac{3}{N_c}) \\[0.5cm]
\displaystyle Z^{DR,\MSbar}_{1,5}  =
\gtilde\,\frac{1}{\epsilon}\,(\frac{2\,N_c}{3}-\frac{3}{N_c}) 
& \displaystyle
Z^{DR,\MSbar}_{1,7}  = -Z^{DR,\MSbar}_{1,9} =
\gtilde\,\frac{1}{\epsilon}\,(-\frac{3\,N_c}{4} +
\frac{3}{2\,N_c}) \\ [0.5cm]
\displaystyle Z^{DR,\MSbar}_{1,3} = Z^{DR,\MSbar}_{1,4} = Z^{DR,\MSbar}_{1,6} =
Z^{DR,\MSbar}_{1,8} = 0 \hspace{-3cm}
\end{array}
\eea

An immediate, well-known by-product of $Z^{DR,\MSbar}_{1,1}$ is the 
anomalous dimension $\widetilde\gamma_{CM}$ for the operator
  $\widetilde{\cal O}_{CM}$\,: $\widetilde\gamma_{CM}  =
g^2/(16\,\pi^2)\cdot (4\,N_c -8/N_c)$.
We note that the mixing coefficients for the gauge noninvariant
operators ${\cal O}_{9}$, ${\cal O}_{10}$ do not vanish.

The ${\cal O}(\epsilon^0)$ parts of the right-hand side of
Eqs.~(\ref{GF}) are the $\MSbar$-renormalized Green's functions; while
they were not necessary for $Z^{DR,\MSbar}_{1i}$, we do 
need them for $Z^{L,\MSbar}_{1i}$ below.

\subsection{Lattice Regularization and $\MSbar$ Renormalization}

The relations which we must turn to now are formally the same as the
ones we studied in the previous subsection (Eqs.~(\ref{GF})); however
all renormalization functions $Z$ now stand for $Z^{L,\MSbar}$, and
the bare Green's functions on the right-hand sides must be calculated
using the lattice 
regularization. The $\MSbar$-renormalized Green's functions on the left-hand
sides coincide with those which we calculated in the previous
subsection, since they must be regularization-independent.

Renormalizability implies that, modulo terms which vanish as
$\Red{a}\to 0$, $\langle \psi^R\,{\cal
    O}_1^R\,\bpsi^R \rangle_{{\rm amp}}
 -\langle\psi\,{\cal O}_1\,\bpsi\rangle _{{\rm amp}}$ 
is polynomial in $m$'s, $q$'s (of $2^{\rm nd}$ degree, but also
$\Red{a^{-1}\cdot}(1^{\rm st})$, $\Red{a^{-2}\cdot}(0^{\rm th})$):
\bea &&\hspace{-1.3cm}\langle \psi^R\,{\cal O}_1^R\,\bpsi^R \rangle_{{\rm amp}}
 -\langle\psi\,{\cal O}_1\,\bpsi\rangle _{{\rm amp}}=
\rho_1\,(q^2_s {+} q^2_d) + \rho_2\,(m_s^2 {+} m^2_d) +
\rho_3\,i\,(m_d\qqq {+} m_s\qq)  
+ \rho_4\,i\,(m_s\qqq {+} m_d\qq)\hspace{-3cm}\nonumber\\
&& \hspace{-0.3cm} + \rho_5\,q_s{\cdot}q_d +
\rho_6\,\qq\,\qqq + \rho_7\,m_sm_d 
+ \Red{{\bf\rho_8}}\,(r_d\,\gamma_5\,\qqq {+} r_s\,\qq\,\gamma_5) 
+ \Red{{\bf\rho_9}}\,i\,(r_dm_d {+} r_sm_s)\,\gamma_5 +
\Red{{\bf\rho_{10}}}\cdot 1 
\label{2ptdiff}\eea
Similarly, $\langle \psi^R\,{\cal
    O}_1^R\,\bpsi^R A_\nu^R \rangle_{{\rm amp}} - \langle\psi\,{\cal
  O}_1\,\bpsi\,A_\nu \rangle _{{\rm amp}}$ 
must be polynomial ($1^{\rm st}$
degree, but also $\Red{a^{-1}\cdot}(0^{\rm th})$):
\bea\langle \psi^R\,{\cal O}_1^R\,\bpsi^R A_\nu^R \rangle_{{\rm amp}} 
- \langle\psi\,{\cal O}_1\,\bpsi\,A_\nu \rangle _{{\rm amp}} &=&
R_1\,g
\,(q_s{+}q_d)_\nu + R_2\,g\,(\gamma_\nu\qqq+ \qq\gamma_\nu)
+ R_3\,i\,g\,(m_s{+}m_d)\gamma_\nu \nonumber \\
&&+ R_4\,(-2i\,g\,\sigma_{\rho\nu}q_{A\rho}) 
+ \Red{{\bf R_5}}\,g\,(r_d - r_s)\,\gamma_5\,\gamma_\nu 
\label{3ptdiff}\eea
The 10+5 coefficients $\rho_i$, $R_i$ depend on
  $\Red{a}$ as: $\Red{a}^{-2}$, $\Red{a}^{-1}$,
$\log(\Red{a}\,\bar\mu)$ (the scale $\bar\mu$
appears through the $\MSbar$-renormalized Green's functions);
they also depend on: $N_c$\,, $\alpha$, and the Symanzik coefficients.

Thus, enforcing Eqs.~(\ref{GF}) leads to 10+5 constraints for
the 13 functions
$Z_{1i}$\,, in such a way as to absorb the above polynomial differences. 
These constraints are self-consistent and complete.

It is a highly nontrivial task to show that the left-hand sides of
Eqs.~(\ref{2ptdiff}, \ref{3ptdiff}) are polynomial. This is especially
true for Eq.~(\ref{3ptdiff}): The $\MSbar$-renormalized 3-point
function has already an extremely complicated
dependence on momenta and masses (involving Spence
functions even for $m=0$), while the lattice bare 3-point function
contains $\sim 10^5$ loop integrals depending on masses and external
momenta. One relatively easy way to show the above property is to make
a special \Red{nondegenerate} choice for the external momenta,
e.g. the ``democratic'' one:
$q_s {-} q_d {+} q_A {=} 0,\  q_s^2 {=} q_d^2 {=} q_A^2 {=} \bar\mu^2$.
In order to subject our results to as stringent a test as possible, we
showed polynomiality without making any simplifying assumptions
on the values of the external momentum 4-vectors, not even momentum
conservation. An independent (far simpler) check on the lattice
Green's functions is that the coefficients of $\log(\Red{a})$ must
match those of $-1/(2\epsilon)$ in DR.

Solving the constraint equations we find, in the case of the tree-level
Symanzik gluon action:
\bea
Z^{L,\MSbar}_{1,1} &=& 1+ \frac{g^2}{16\pi^2}\,
    \bigl(N_c (-12.8455 +\frac{1}{2}\,\log(a^2\,\bar\mu^2 ))
    + \frac{1}{N_c}  (9.3779
    -\frac{5}{2}\,\log(a^2\,\bar\mu^2 ) ) \bigr), 
\\
Z^{L,\MSbar}_{1,2} &=& \frac{g^2C_F}{16\pi^2}\,(2.7677 + 6\,\log(a^2\,\bar\mu^2 )),
\ \ 
Z^{L,\MSbar}_{1,3}  = 0,
\ \ 
Z^{L,\MSbar}_{1,4}  = 0,
\nonumber\\
Z^{L,\MSbar}_{1,5} &=& \frac{g^2}{16\pi^2}\,
                   \bigl(N_c (5.3894 -\frac{3}{2}\, \log(a^2\,\bar\mu^2 )) + 
                         \frac{1}{N_c}  (-5.5061 + 3\,\log(a^2\,\bar\mu^2 )) \bigr),
\ \ 
Z^{L,\MSbar}_{1,6}  = 0,
\nonumber\\
Z^{L,\MSbar}_{1,7}  &=& -Z^{L,\MSbar}_{1,9} = -\frac{Z^{L,\MSbar}_5}{2}\,,
\ \ 
Z^{L,\MSbar}_{1,8}  = \frac{g^2C_F}{16\pi^2}\,(-3.9654),
\ \  
Z^{L,\MSbar}_{1,10} =\frac{g^2C_F}{16\pi^2}\,(5.5061 - 3\,\log(a^2\,\bar\mu^2 )), 
\hspace{-0.5cm}
\nonumber\\
Z^{L,\MSbar}_{1,11} &=&
\frac{1}{\Red{a}}\,\frac{g^2C_F}{16\pi^2}\,(-4.0309) = -Z^{L,\MSbar}_{1,12} ,
\ \ 
Z^{L,\MSbar}_{1,13} = \frac{1}{\Red{a}^2}\frac{g^2C_F}{16\pi^2}\,(47.7929) 
\nonumber
\eea
Systematic errors coming from numerical loop integration are much smaller than the precision
presented in the above results. Also, certain mixing coefficients
vanish at one loop, but not beyond.

\subsection{Non-perturbative results -- Preliminary}

In the calculation of on-shell matrix elements, by virtue of the
equations of motion, some of the operators 
${\cal O}_1$ -- ${\cal O}_{13}$ will not appear. The remaining ones: ${\cal O}_1$\,,
${\cal O}_2$\,, ${\cal O}_3$\,,  ${\cal O}_4$\,,  ${\cal O}_{12}$\,,  ${\cal
  O}_{13}$\, will be present, and it is imperative to have a stringent
estimate of the corresponding mixing coefficients. For operators
of the same dimensionality as the chromomagnetic one, i.e. ${\cal O}_1$\,,  ${\cal O}_2$\,,
${\cal O}_3$\,,  ${\cal O}_4$\,, our one-loop results are expected to
provide satisfactory accuracy; however, for operators of lower
dimensionality (${\cal O}_{12}$\,,  ${\cal O}_{13}$), given that their
coefficients are power divergent, perturbation theory is expected
to provide only a ballpark estimate at best. Fortunately, it is
precisely for the coefficients of these latter operators that we can have best access to
non-perturbative estimates.

Imposing conditions such as:
\be\lim_{m_s\,,\ m_d \to 0} \langle\pi(0)|{\cal O}_1^{\rm
  sub}|K(0)\rangle = \lim_{m_s\,,\ m_d \to 0} \langle\pi(0)|{\cal O}_1
  + \frac{\NavyBlue{c_{13}}}{\Red{a^2}}{\cal O}_{13}|K(0)\rangle = 0\ee
\vspace{-0.3cm}
\be\langle 0|{\cal O}_1^{\rm sub}|K(0)\rangle_{m_s\,,\ m_d} = \langle 0|{\cal O}_1
  + \frac{\NavyBlue{c_{13}}}{\Red{a^2}}{\cal O}_{13} 
+ \frac{\NavyBlue{c_{12}}}{\Red{a}}{\cal O}_{12}|K(0)\rangle_{m_s\,,\ m_d} = 0\ee
we can fit the values of $c_{13}(g_0)$, $c_{12}(g_0)$ to data from simulations with
varying quark masses. 

In a preliminary series of simulations~\cite{CMO}, we have extracted
 $c_{13}$ at different values of the coupling
($\beta\equiv 6/g_0^2 = 1.90,\ 1.95,\ 2.10$). The results for $c_{13}$
closely follow a quadratic dependence on $g_0$\,, thus resembling a
one-loop effect; nevertheless there is a notable difference, as was
expected:
\be
Z_{1,13}^{\rm non-pert} \sim \Red{a^{{-}2}} \,\frac{g^2C_F}{16\pi^2} \,(33.7)\qquad\qquad
Z_{1,13}^{\rm pert} = \Red{a^{{-}2}}\,\frac{g^2C_F}{16\pi^2} \,(47.793)
\ee
[For a discussion on the possible choices for the coupling constant,
  see Ref.~\cite{CMO}.]

\section{Checks -- Extensions}
Besides a series of controls which we have applied to our results,
some further ones may be applied: (i) A calculation of 
4-point Green's functions will provide important consistency checks,
but no new information, on $Z_{1i}$\,. On the other hand, 5-point functions and
beyond are irrelevant: Being superficially convergent, they have a
straightforward continuum limit. (ii)
Non-perturbative estimates of all mixing coefficients would be
  very important cross checks.

Depending on the method one wishes to employ for computing
matrix elements of the CMO non-perturbatively, a renormalization
scheme other than $\MSbar$ may be more appropriate. In particular, one
may employ an extension of the RI' scheme, in which RI'-like
conditions need to be imposed on both 2-point and 3-point functions.
The new mixing coefficients $Z^{L,RI'}_{ij}$ are
related to $Z^{L,\MSbar}_{ij}$ via a ($13\times 13$) regularization-independent
conversion matrix, whose elements are finite functions of the
renormalized coupling. In fact, all relevant matrix elements are
directly obtainable from our results on the renormalized Green's
functions, with no further calculation required.

A further extension of the present work would be to apply methods of
improved perturbation theory (``boosted'' coupling, ``cactus''
diagrams, etc.) to our results. Another direction is to compute
${\cal{O}}(a^2 g^2)$ corrections to Green's functions; these, combined
with non-perturbative evaluations, lead to an improvement in the
non-perturbative estimates of the mixing coefficients.

\bigskip\noindent
{\bf Acknowledgments:} The work of M.~Constantinou and M.~Costa was
supported by the Cyprus Research Promotion Foundation under Contract
No. TECHNOLOGY/${\rm \Theta E\Pi I\Sigma}$/0311(BE)/16. D.~Meloni
acknowledges MIUR (Italy) for financial support under the program
Futuro in Ricerca 2010 (RBFR10O36O).

\end{document}